\RequirePackage{ifpdf}
\ifpdf 
\documentclass[pdftex]{sigma}
\else
\documentclass{sigma}
\fi

\begin{document}

\numberwithin{equation}{section}

\allowdisplaybreaks

\renewcommand{\PaperNumber}{021}

\FirstPageHeading

\renewcommand{\thefootnote}{$\star$}

\ShortArticleName{Eigenvalues of Killing Tensors  and Separable
Webs}

\ArticleName{Eigenvalues of Killing Tensors  and Separable Webs\\
on Riemannian and Pseudo-Riemannian Manifolds\footnote{This paper
is a contribution to the Vadim Kuznetsov Memorial Issue
``Integrable Systems and Related Topics''. The full collection is
available at
\href{http://www.emis.de/journals/SIGMA/kuznetsov.html}{http://www.emis.de/journals/SIGMA/kuznetsov.html}}}

\Author{Claudia CHANU and Giovanni RASTELLI}

\AuthorNameForHeading{C. Chanu and G. Rastelli}

\Address{Dipartimento di Matematica, Universit\`a di Torino, Via
Carlo Alberto 10, 10123 Torino, Italy}
\Email{\href{mailto:claudiamaria.chanu@unito.it}{claudiamaria.chanu@unito.it},
\href{mailto:giorast@tin.it}{giorast@tin.it}}

\ArticleDates{Received November 02, 2006, in f\/inal form January
16, 2007; Published online February 06, 2007}

\Abstract{Given a $n$-dimensional Riemannian manifold of arbitrary
signature, we illustrate an algebraic method for constructing the
coordinate webs separating the geodesic Hamilton--Jacobi equation
by means of the eigenvalues of $m\leq n$ Killing two-tensors.
Moreover, from the analysis of the eigenvalues, information about
the possible symmetries of the web foliations arises. Three cases
are examined: the orthogonal separation, the general separation,
including non-orthogonal and isotropic coordinates, and the
conformal separation, where Killing tensors are replaced by
conformal Killing tensors. The method is illustrated by several
examples and an application to the L-systems is provided.}


\Classification{70H20; 70G45}

\section{Introduction}

It is well-known that the separation of variables in the
Hamilton--Jacobi equation of a natural Hamiltonian is
characterized by the existence of sets of Killing 2-tensors
($\mathcal K_m$) and Killing vectors~($D_r$) with suitable
properties \cite{[13],[4],[6]}. The aim of this paper is to show
some geometrical properties of a separable web (i.e., a set of
foliations of coordinate hypersurfaces) by means of the analysis
of $D_r$ and $\mathcal K_m$; in particular, we provide an
algebraic method for determining the equations of the separable
coordinate hypersurfaces. For instance, it is known that on a
two-dimensional manifold the separation  is characterized by a
Killing tensor $\mathbf K$ with pointwise simple eigenvalues
$(\lambda^1,\lambda^2)$. Separable coordinates $(q^1,q^2)$ can be
determined in two ways: (i) by setting $q^1=\lambda^2$ and
$q^2=\lambda^1$ or, (ii) by integrating the eigenvectors $(\mathbf
E_1,\mathbf E_2)$, which are orthogonal to the coordinate curves.
In case (i) we need to assume that both eigenvalues are real
independent functions (except for a closed subset of the
conf\/iguration manifold at most); if the eigenvalues are not
independent (for instance one of them is constant), then
symmetries i.e., Killing vectors or ignorable coordinates, are
present and the problem becomes even simpler. We remark that
method (i) is purely algebraic: starting from $\mathbf K$, we have
only to solve its characteristic equation.

In \cite{[10]} we extended this analysis, for the orthogonal
separation, to the $n$-dimensional Riemannian manifolds. For $n>2$
the eigenvalues of the Killing tensors do not def\/ine directly
separable coordinates, however, we show how to construct rational
functions of them which are constant on the separable coordinate
hypersurfaces.

In the present paper we analyze the most general case of
separation on a (pseudo) Riemannian manifold, without assumptions
on the signature of the metric and on the orthogonality of the
separable coordinates. In the following we shall refer to this
kind of separation as ``non-orthogonal separation" or ``general
separation". In non-orthogonal separation a fundamental role is
played by a particular $m$-dimensional space of Killing tensor
$\mathcal K_m$ ($m \leq n$), whose properties relevant for our
aims are recalled in Section \ref{s1}. We illustrate an algebraic
algorithmic method for constructing a set of intrinsically
def\/ined functions, that we call {\it fundamental functions},
 starting from a special class of eigenvalues of $m$ independent Killing tensors of $\mathcal K_m$.
The analysis of these functions allows us to detect two classes of
symmetries ({\it proper} and {\it conformal}) of the St\"ackel
matrix associated with separation. When no proper symmetries
occur, the fundamental functions are an ef\/fective tool for
computing the equations of the separable coordinate hypersurfaces.
Other algorithmic approaches for f\/inding separable coordinates
are already known in important particular cases, such as
orthogonal separation in constant curvature manifolds with
positive def\/inite metric \cite{[18],rauch}, essentially based on
the previous knowledge of all separable coordinates in these
spaces (as generalized elliptic coordinates and their
degenerations).

A further generalization considered here is the analysis  of the
case of the so-called {\it orthogonal conformal separation}, which
deals with separable coordinates for the null-geodesic
HJ-equation, or for a HJ-equation of a natural Hamiltonian with
f\/ixed value of the energy \cite{[7]}. The orthogonal separable
coordinates can be considered as a special case of this broader
class of coordinates. In the intrinsic characterization of the
orthogonal conformal separation, the Killing tensors are replaced
by conformal Killing tensors with suitable properties. Following
the same procedure of the ``ordinary" separation, we are able to
construct intrinsic functions allowing to deduce geometrical
properties of the conformal separable coordinate hypersurfaces or
to construct their equations.

\looseness=1 In Section \ref{s1} we recall the basic intrinsic
characterizations of the non-orthog\-onal separation on a
Riemannian manifold in a form suitable for our needs. In Section
\ref{s2}  we describe our method and its application to a simple
example. In Section \ref{s3}, devoted to the orthogonal
separation, we improve the analysis given in \cite{[10]} and we
show the links between eigenvalues of Killing tensors and  proper
or conformal symmetries  for the associated coordinate systems. In
the orthogonal case, by ``proper symmetry" (resp., ``conformal
symmetry") of a coordinate system we mean that there are Killing
vectors  (resp.,  conformal Killing vectors) orthogonal to some
foliations of the web. In Section \ref{s4}, we see how the
def\/initions of proper and conformal symmetries of the
coordinates can be extended for the general separation and
 we generalize our results for the cases when non-orthogonal or null coordinates occur.
In Section \ref{s5}, we summarize the intrinsic characterization
of the orthogonal conformal separation \cite{[7],[14]} and we
apply our algebraic method to the case of  conformal separable
orthogonal webs, showing how to detect conformal symmetries and to
write the equation describing the foliations without conformal
symmetries.

Each section is completed by  illustrative examples: the spherical
coordinates in ${\mathbb R}^3$ (Section~\ref{s2}), the L-systems
\cite{[5]}, also known as Benenti systems \cite{[8],[9],tsiga}
(Section~\ref{s3}), two non-orthogonal 4-dimensional  coordinate
systems (one of them with null coordinates)  in Section \ref{s4},
and the conformal separable coordinate system, known as
tangent-spheres coordinates \cite{[15]} (Section \ref{s5}).
Moreover, by applying our analysis to  L-systems, we prove an
interesting geometrical property of these systems: for $n>2$, none
of the common eigenvectors of the associated Killing tensors is a
proper conformal Killing vector.

\section{An outline of geodesic separation on Riemannian\\ and pseudo-Riemannian manifolds}\label{s1}

We consider a $n$-dimensional Riemannian manifold $Q$ with
contravariant metric $\mathbf G=(g^{ij})$ of arbitrary signature
and the corresponding {\it geodesic Hamiltonian} $G=\frac 12
g^{ij}p_ip_j$. A relation of equivalence is def\/ined among
separable coordinate systems for $G$ \cite{[1],[2]} such that in
each class there are the particular coordinate systems described
in Theorem \ref{t1.1} below. We recall that a~regular square
matrix $(\varphi_j^{(i)})$ is a {\it St\"ackel matrix} if any
element $\varphi_j^{(i)}$ is a function of the coordinate~$q^j$
only.

\begin{theorem}[\cite{[1],[2]}] \label{t1.1}
In an equivalence class of separable coordinates there exists a
{\rm standard coordinate system} $(q^{\hat a},q^{\bar a},
q^{\alpha})$ such that {\rm (i)} The metric tensor has the {\rm
standard form}
   \begin{equation} \label{1.1}
(g^{ij})=\bordermatrix{
&\overbrace{\hphantom{AAA}}^{\displaystyle{m_1}}
&\overbrace{\hphantom{AAA}}^{\displaystyle{m_0}}
&\overbrace{\hphantom{AAA}}^{\displaystyle{r}}\cr
m_1\phantom{r}\bigg\{   & g^{\hat a\hat a} & 0 & 0 \cr
m_0\phantom{r}\bigg\{   & 0 & 0 & g^{\bar b\beta} \cr
r\phantom{m_0}\bigg\{   & 0 & g^{\alpha\bar a} &
g^{\alpha\beta}\cr},
  \end{equation}
where the coordinates $(q^{\alpha})$ $(\alpha=m_0+m_1+1,\ldots,n)$
are  {\sl ignorable}, $\partial _{\alpha}g^{ij}=0$
$(i,j=1,\ldots,n)$. {\rm(ii)} The non-vanishing metric components
have the form
  \begin{equation}\label{1.2}
\begin{array}{l}
g^{\hat a\hat a}=\varphi^{\hat a}_{(m)},\\
g^{\bar a \beta}=\theta^{\beta}_{\bar a}\varphi^{\bar a}_{(m)},\\
g^{\alpha \beta}=\eta^{\alpha \beta}_a\varphi^a_{(m)},
\end{array}
\qquad \begin{array}{l}
 a=1,\ldots, m_1+m_0, \\
 \hat a=1,\ldots, m_1, \\
 \bar a=m_1+1,\ldots, m_1+m_0, \\
 \alpha,\beta =m_1+m_0+1,\ldots, n ,
\end{array}
  \end{equation}
where $\theta^\beta_{\bar a}$ and $\eta^{\alpha\beta}_a$ are
functions of the coordinate corresponding to the lower index only
and $(\varphi ^a_{(m)})=(\varphi^{\hat a}_{(m)},\varphi^{\bar
a}_{(m)})$, $m=m_0+m_1$,  is a row of the inverse of a $m\times m$
St\"ackel matrix $\mathbf S$ in the coordinates $(q^a)$.
\end{theorem}


\begin{theorem}[\cite{[4]}] \label{t1.2}
The geodesic Hamiltonian $G$ is separable if and only if there
exists a pair $(D_r,\mathcal K_m)$, called {\sl separable Killing
algebra}, such that: a) $D_r$ is a r-dimensional Abelian algebra
of Killing vectors spanning a regular distribution $\Delta$ of
rank $r$ such that $I=\Delta \cap \Delta^{\perp}$ has constant
rank $m_0$. b) $\mathcal K_m$ is a m-dimensional space of Killing
tensors, generated by  $m$ independent tensors $(\mathbf K_a)$,
with $m$ common eigenvectors orthogonal to $D_r$ which are normal
(i.e., orthogonally integrable) and associated with real
eigenvalues. c) $\mathcal K_m$ is $D_r$-invariant. d) For $m_0>1$,
$d(\mathbf K\cdot dg^{\alpha \beta})=0$ for any $\mathbf K\in
\mathcal K_m$, where $g^{\alpha\beta}={\mathbf X}_\alpha \cdot
{\mathbf X}_\beta$ for any basis $({\mathbf X}_\alpha)$ of $D_r$.
\end{theorem}


By ``Killing tensor" (KT) we mean a symmetric two-tensor $\mathbf
K$ such that $[\mathbf K,\mathbf G]=0,$ where $[\cdot,\cdot]$
denotes the Lie--Schouten brackets. With a separable Killing
algebra we associate an important kind of coordinates in the
following way:


\begin{definition}\label{d1.3}
\rm Let $(D_r,\mathcal K_m)$ be a separable Killing algebra and
$(\mathbf E_a)$ be the $m$ normal common eigenvectors $(\mathbf
E_a)$ of the elements  of $\mathcal K_m$. The coordinates
$(q^a,q^\alpha)$ are {\it adapted coordinates} of $(D_r,\mathcal
K_m)$ if the coordinate hypersurfaces of $q^a$ are the integral
manifolds orthogonal to each~$\mathbf E_a$ (or equivalently, the
dif\/ferentials $dq^a$ are common eigenforms of $\mathcal K_m$)
and the vector f\/ields $\partial_\alpha$ form a basis of $D_r$
(i.e., the $q^\alpha$ are the af\/f\/ine parameters of a given
basis $\mathbf X_\alpha$ of $D_r$ with zero values on a chosen
integral manifold of $\Delta^\perp$); the $m$ orthogonal
coordinates $(q^a)$ are said to be {\it essential}. The $m_0$
essential coordinates $(q^{\bar a})$ such that $g^{\bar a \bar
a}=0$ are called {\it isotropic} or {\it null coordinates}.
\end{definition}

The integrability of the distribution $\Delta^\perp$ is a
consequence of the hypotheses of Theorem \ref{t1.2} \cite{[4]}. We
remark that if $Q$ is a proper Riemannian manifold (or if the
coordinates are orthogonal) there are no isotropic coordinates.

\begin{theorem}[\cite{[4]}] \label{t1.4}
Let $(q^i)=(q^{\hat a}\!,q^{\bar a}\!,q^\alpha)$ be an adapted
coordinate system of a separable Killing algebra $(D_r,\mathcal
K_m)$. Then, {\rm (i)} The coordinates $(q^i)$  are standard
separable coordinates and each tensor $\mathbf K\in\mathcal K_m$
assumes the standard form
  \begin{equation} \label{1.3}
(K^{ij})=
\begin{pmatrix}
\lambda^{\hat a}\,g^{\hat a\hat a}  &   0   &   0   \cr &&\cr 0
&   0   &   \lambda^{\bar b}\,g^{\bar b\beta} \cr &&\cr 0   &
\lambda^{\bar a}\,g^{\alpha\bar a} & K^{\alpha\beta}\cr
\end{pmatrix}.
  \end{equation}
{\rm (ii)} Given a basis $(\mathbf K_{a})$ of $\mathcal K_m$, the
non-vanishing components of each $\mathbf K_{b}$ $(b=1,\ldots, m)$
assume the form
  \begin{equation} \label{1.4}
\begin{array}{l}
K_b^{\hat a\hat a}=\varphi^{\hat a}_{(b)},\\
K_b^{\bar a \beta}=\theta^{\beta}_{\bar a}\varphi^{\bar a}_{(b)},\\
K_b^{\alpha \beta}=\eta^{\alpha \beta}_a\varphi^a_{(b)},
\end{array}
\qquad \begin{array}{l}
 a=1,\ldots, m, \quad
 b=1,\ldots, m,\\
 \hat a=1,\ldots, m_1, \quad \bar a=m_1+1,\ldots, m, \\
 \alpha,\beta =m+1,\ldots, n,
\end{array}
  \end{equation}
where $(\varphi^a_{(b)})$ is a row of $\mathbf S^{-1}$.
 \end{theorem}


\begin{remark}\label{r1.5} \rm  The functions $(\lambda^a)$ are the eigenvalues of $\mathbf K$ corresponding to the common eigenvectors of all tensors in $\mathcal K_m$ and satisfy  the following  {\sl intrinsic Killing equations}:
  \begin{equation} \label{1.5}
\begin{array}{l}
\partial _a\lambda ^b=(\lambda^a-\lambda^b)\partial _a\ln \varphi^b_{(m)}, \\
\partial _aK^{\alpha \beta}=\lambda^a\partial_ag^{\alpha \beta}, \\
\partial _{\alpha}\lambda^j=0,
\end{array}
\qquad a=(\hat a,\bar a),\quad a,b=1,\dots, m,\quad j=1,\dots, n.
  \end{equation}
\end{remark}

The geometric realization of an equivalence class of separable
coordinates  is called {\it separable Killing web} \cite{[4]}:


\begin{definition} \label{d1.6} \rm
A {\it separable Killing web} is a triple $(\mathcal S_m,
D_r,\mathbf K)$, where (i) $\mathcal S_m=(S^a)$ is a set of $m$
foliations of hypersurfaces pairwise transversal and orthogonal;
(ii) $D_r$ is a $r$-dimensional Abelian algebra of Killing vectors
tangent to each foliation $S^a$; (iii) $\mathbf K$ is a
$D_r$-invariant Killing tensor with $m$ eigenvectors $(\mathbf
E_a)$ associated with $m$ pointwise distinct real eigenvalues
$(\lambda^a)$ and orthogonal to the foliations $(S^a)$. The KT
$\mathbf K$ is called {\it characteristic tensor of the web}.
\end{definition}

\begin{remark}\label{d1.7} \rm
The existence of a separable Killing web is equivalent to the
existence of a separable Killing algebra, or of separable
coordinates for $G$. Indeed, a separable Killing web $(\mathcal
S_m,D_r,\mathbf K)$ gives rise to a standard separable coordinate
system $(q^a,q^\alpha)$ such that the coordinate hypersurfaces
$q^a=\hbox{constant}$ are leaves of $S^a$ and the vector f\/ields
$(\partial_\alpha)$ associated with $q^\alpha$ form a basis of
$D_r$.
\end{remark}

From Def\/initions \ref{d1.3} and \ref{d1.6}, it follows that essential coordinates only are associated with the eigenvectors of Killing tensors, that is with the foliations $(S^a)$ of a separable Killing web. Therefore, in the following sections we restrict our attention to the essential coordinates.
In  Example~\ref{e4.14}  we show a separable Killing web, the
corresponding separable  Killing algebra and the St\"ackel matrix
in the adapted coordinates for the case $n=4$, $m=2$ and $m_0=0$.

\section{The method of the eigenvalues}\label{s2}

In order to clarify the exposition, we collect in this section the
results proved in Sections \ref{s3}--\ref{s5}. We recall that


\begin{definition}\label{d2.1} \rm A vector f\/ield $\mathbf X$ is said to be a {\it conformal Killing vector} (CKV) if there exists a~function $F$
such that
\[
[\mathbf X,\mathbf G]=\mathcal L_{\mathbf X}\, \mathbf G=F\mathbf
G,
\]
 where $[\cdot,\cdot]$ is the Schouten bracket and $\mathcal L$ the Lie-derivative.  If $F=0$, $\mathbf X$ is a Killing vector (KV) and if $F \neq 0$ we call $\mathbf X$ a {\it proper} CKV.
\end{definition}

\begin{remark}\label{r2.2} \rm
A coordinate $q^{i}$ is ignorable if and only if $\partial _{i}$
is a Killing vector. Moreover, in an orthogonal system, if
$\partial _{i}$ is proportional to a KV (i.e., there is a function
$f$ such that $f\partial_i$ is a~KV) then $q^i$ is ignorable up to
a rescaling $\tilde q^i=\tilde q^i(q^i)$.
\end{remark}

\begin{note} \label{n2.3}
Here and in the following, for any matrix $\mathbf A=(a_j^i)$ the
lower index  is the row-index and the upper one is the column
index. Moreover, we shall denote by $\mathbf A_j^i$ the submatrix
of $\mathbf A$ obtained by eliminating the $j$-th row and the
$i$-th column.
\end{note}

\noindent \textsc{Step 1. Construction of the fundamental
functions.}

Let $(D_r,\mathcal K_m)$ be a separable Killing algebra
associated with a separable web $(D_r, \mathcal S_m)$ and let
$(\mathbf K_1,\ldots,$ ${}\mathbf K_m=\mathbf G)$ be a basis of
$\mathcal K_m$.

\begin{itemize}\itemsep=0pt
\leftskip .0cm \item[i)]
 We determine the essential eigenvalues of $(\mathbf K_a)$ associated with the common essential eigenvectors  $(\mathbf E_1,\ldots, \mathbf E_m)$ orthogonal to $D_r$.


\item[ii)] We construct the  regular (see Remark \ref{r:ult})
$m\times m$ matrix $\Lambda=(\lambda_a^b)$ of the essential
eigenvalues of $\mathbf K_a\in\mathcal K_m$ ordered as follows:
$\lambda_a^b$ is the eigenvalue of $\mathbf K_a$ associated with
the common eigenvector $\mathbf E_b$.
\\
We remark that for the construction of $\Lambda$ we have to order
properly the eigenvalues of each KT and, to do that, we need to
compute the eigenvectors. However, our further analysis is based
only upon the matrix $\Lambda$ of the eigenvalues and no
integration is needed.


\item[iii)] For $a,b,c=1,\ldots,m$ we consider the intrinsic
ratios
  \begin{equation} \label{2.1}
f_a^{bc}=\frac{\det \Lambda_b^a}{\det \Lambda_c^a},
  \end{equation}
that we call {\it fundamental functions}. They are  {\it
well-defined} functions only if $\det \Lambda_c^a$  is not
identically zero. Moreover, since $f_a^{bc}=1/f_a^{cb}$ and
$f_a^{bb}$ is equal to 1 or everywhere undef\/ined, in the
following we shall assume $b>c$.

\end{itemize}

\noindent\textsc{Step 2. Analysis.}

Let us f\/ix an index $a$. Due to the regularity of $\Lambda$, at
least one fundamental function (\ref{2.1}) is well def\/ined
(Proposition \ref{p3.3}). By examining the functions $f_a^{bc}$
written in an arbitrary coordinate system, we can easily detect
symmetries of the foliation $S^a\in \mathcal S_m$;  if $S^a$ has
no symmetries we obtain the equation of the foliation $S^a$.
Indeed, two dif\/ferent and mutually exclusive cases occur:

\begin{itemize}\itemsep=0pt
\item[iv)]
 There exist indices $c$ and $b$ such that the function $f_a^{bc}$ is not identically constant. In this case, in a neighborhood of each point $P_0$ such that $df_a^{bc}(P_0)\neq 0$, equation
\[
f_a^{bc}=f_a^{bc}(P_0)
\]
def\/ines a  hypersurface containing $P_0$ and orthogonal to the
eigenvector $\mathbf E_a$ (i.e., a hypersurface of $S^a$). Hence,
equations
\[
f_a^{bc}=k,
\]
for suitable values of $k\in {\mathbb R}$, describe the foliation
$S^a$ (see Theorems \ref{t3.8} and \ref{t4.11}, for the orthogonal
and the general case respectively).


\item[v)] For the f\/ixed index $a$ all functions $f_a^{bc}$
constructed in \textsc{Step 1} are constant or undef\/ined. Then,
special properties of the adapted coordinates of $(D_r,\mathcal
K_m)$
hold. Let $(q^1,\ldots,q^m)$ be the essential coordinates adapted
to the foliations $(S^1,\ldots,S^m)$. Up to a reparameterization
of $q^a$, the St\"ackel matrix $\mathbf S=(\varphi_c^{(b)})$ and
its inverse matrix $\mathbf
S^{-1}=(\varphi_{(c)}^b)=(\lambda_c^b\varphi_{(m)}^b)$ do not
depend on $q^a$ (see Theorem \ref{t4.9} (ii)). We call the vector
f\/ield $\partial_a$ a {\it St\"ackel symmetry} (see Section
\ref{s4} for further details).
\end{itemize}

We remark that we do not need to distinguish between isotropic and
non-isotropic coordinates.


Moreover, if we consider the orthogonal separation (i.e., when
$m=n$, $D_r=0$ and all coordinates are essential) then we are able
to test if a foliation $S^a$ is orthogonal to a Killing vector or
to a conformal Killing vector, by examining the fundamental
functions. Indeed, the following properties hold:
\begin{itemize}\itemsep=0pt
\item[vi)] All fundamental functions $f_a^{bc}$  are constant or
undef\/ined if and only if the eigenvector $\mathbf E_a$ is
proportional to a Killing vector i.e., the associated adapted
coordinate $q^a$ is (up to a~reparameterization) ignorable
(Theorem \ref{t3.6} (ii)).
\item[vii)] All fundamental functions $f_a^{bc}$ with $c<b<n$
($n>2$) are constant or undef\/ined if and only if the
corresponding eigenvector $\mathbf E_a$ is proportional to a
conformal Killing vector (Theorem~\ref{t3.6}~(i)).
\end{itemize}

\begin{remark} \label{r2.4} \rm
Also for the general separation, properties analogous to items vi)
and vii) hold. Item vi) is in fact a special case of the general
situation described in item v), holding for orthogonal
coordinates. Indeed, due to Remark \ref{r2.2} and to equations
(\ref{1.5})$_1$, we have that $\mathbf E_a$ is proportional to a
Killing vector if and only if the St\"ackel matrix and its inverse
do not depend on the corresponding coordinate $q^a$. The property
stated in item vii) can be extended to general separable
coordinates as follows (Theorem \ref{t4.9} (i)):
All functions $f_a^{bc}$ with $c<b<m$ ($m>2$) are constant or
undef\/ined if and only if there exists a function $F$ such that
(up to a~reparameterization of $q^a$) $\partial_a
\varphi_{(m)}^b=F\varphi_{(m)}^b$ for all $b=1,\ldots,m$.
Then, we call $\partial_a$ a {\it conformal St\"ackel symmetry}
(see Section \ref{s4} for further details).
\end{remark}

\begin{remark} \label{r2.5} \rm
For $m=2$,  we have only two fundamental functions:
$f_1^{21}=\lambda^2$  and $f_2^{21}=\lambda^1$ i.e., the
eigenvalues of the characteristic tensor.
\end{remark}

We show how the method works in the following simple but
illustrative example.

\begin{example} \label{e2.6} \rm
Let us consider in ${\mathbb R}^3$ the spherical coordinates
centered at a point $O$, with axis $\omega$ passing through $O$
and parallel to a unit vector $\mathbf n$. It is well known that
they are orthogonal separable coordinates for the geodesic
Hamiltonian. Thus, we have $m=n=3$ and all coordinates are
essential. A basis of  $\mathcal K_3$ is
\[
\mathbf K_1=r^2\mathbf G-\mathbf r\otimes \mathbf r, \qquad
\mathbf K_2=(\mathbf n\times\mathbf r)\otimes (\mathbf n\times
\mathbf r),\qquad \mathbf K_3=\mathbf G,
\]
where $\mathbf r$ is the position vector with respect to $O$ and
$r=\| \mathbf r\|$. The common eigenvectors are
\[
\mathbf E_1=\mathbf r, \qquad \mathbf E_2=\mathbf r\times (\mathbf
n\times \mathbf r),\qquad \mathbf E_3=\mathbf n\times \mathbf r,
\]
which are orthogonal to the foliation $S^1$ of the spheres
centered at $O$, to the foliation $S^2$ of the circular cones with
vertex $O$ and axis $\mathbf n$ and to the foliation $S^3$ of the
meridian half-planes  issued from $\omega$, respectively. The
matrix $\Lambda$ of the eigenvalues of $\mathbf K_a$ is
\[
\Lambda=\begin{pmatrix}
   0 & r^2 & r^2 \\
   0 & 0 & \|\mathbf n\times \mathbf r\|^2 \\
   1 & 1 & 1 \end{pmatrix}.
\]
By computing  $\det \Lambda^a_b$ for $a,b=1,\ldots,3$, we get that
the non vanishing ones are
\[
\Lambda_1^1=\Lambda_1^2=-\|\mathbf n\times \mathbf r\|^2, \qquad
\Lambda_3^1=r^2\|\mathbf n\times \mathbf r\|^2, \qquad
\Lambda_2^2=\Lambda_2^3=- r^2.
\]
The fundamental functions (\ref{2.1}) are summarized in the
following table (n.d.\ means that the denominator vanishes
identically and the function is not def\/ined)

\begin{center}
\begin{tabular}[c]{|c|c|c|c|}
\hline
      & $c=1$, $b=2$ &  $c=1$,  $b=3$ &  $c=2$,  $b=3 \vphantom{\dfrac 12}$  \\
\hline
$a=1$ & $f_1^{21}=0$  & $f_1^{31}=-r^2$ & $f_1^{32}{\hbox { n.d.}}\vphantom{\dfrac 12}$  \\ 
\hline
$a=2$ & $f_2^{21}=\frac{r^2}{\parallel \mathbf n\times\mathbf r\parallel^2}$ & $f_2^{31}=0$ & $f_2^{32}=0\vphantom{\dfrac 12}$  \\ 
\hline
$a=3$ & $f_3^{21} {\hbox { n.d.}}$ & $f_3^{31}=0$  & $f_3^{32}{\hbox { n.d.}} \vphantom{\dfrac 12}$ \\
\hline
\end{tabular}
\end{center}

For $a=1$, the function $f_1^{31}=-r^2$ is constant on the
hypersurfaces of $S^1$ and equation $f_1^{31}=k$, for real
negative values of $k$, describes all the spheres of $S^1$.
According to the fact that $\mathbf E_1=\mathbf r$ is a CKV
$(\mathcal L_{\mathbf r}\mathbf G=-2\mathbf G)$, we have that for
all $c<b<3$ all functions $f_1^{bc}$ are constant or undef\/ined.

For $a=2$, the level sets of the non-constant function
$f^{21}_2={r^2}{ \|\mathbf n\times \mathbf r\|^{-2}}$ are the
surfaces of~$S^2$ and, since the upper indices are both $<3$, the
eigenvector $\mathbf E_2$ is not proportional to a~CKV.

For $a=3$, since for any $b$ and $c$, $f_3^{bc}$ is undef\/ined or
identically constant, we have that $\mathbf E_3$~is a Killing
vector (the rotation around the axis $\omega$) and the
corresponding coordinate $q^3$ (the rotational angle) is
ignorable.
\end{example}

\section{Orthogonal separable webs}\label{s3}

We consider a $n$-dimensional Riemannian manifold $Q$ with
positive def\/inite metric $\mathbf G$ and the corresponding
geodesic Hamiltonian $G=\frac 12 g^{ij}p_ip_j$. We suppose that
$G$ is orthogonally separable. We adapt the general results of
Section \ref{s1} to the case $m=n$, $m_0=0$. Thus, some of the
geometric structures introduced in Section \ref{s1} are
simplif\/ied. The separable Killing web (Def\/inition \ref{d1.6})
is replaced by  the {\it orthogonal  separable
 web} $(\mathcal S_n,\mathbf K)$, that is a set of $n$ pairwise orthogonal foliations
 $\mathcal S_n=(S^i)$ orthogonal to the eigenvectors of the Killing tensor $\mathbf K$
with simple eigenvalues (the {\it characteristic tensor} of the
web). In the orthogonal context, the linear space $D_r$ of Killing
vectors disappears and
 the $n$-dimensional space of Killing tensors $\mathcal K_n$ associated with
$\mathcal S_n$ is called {\it Killing--St\"ackel algebra}
(KS-algebra) or {\it Killing--St\"ackel space}. All Killing
tensors of $\mathcal K_n$ have common normal (i.e., orthogonally
integrable) eigenvectors $(\mathbf E_i)$, the integral manifolds
orthogonal to $\mathbf E_i$ are the leaves of $S^i$ and all
coordinates are essential. We denote  by $(\mathbf K_{j})$ a~basis
of $\mathcal K_n$ with $\mathbf K_{n}=\mathbf G$. Adapting Theorem
\ref{t1.4} to the orthogonal separation, we get

\begin{proposition}\label {p3.1}
Let $(q^i)$ be a coordinate system adapted to the KS-algebra
$\mathcal K_n$. Then, {\rm (i)}
 the $(q^i)$ are orthogonal separable coordinates for $G$.
$(ii)$ Given a basis $(\mathbf K_{j})$ of $\mathcal K_n$, we have
\[
\mathbf K_{j}=\sum_i \lambda_{j}^i g^{ii}\, \mathbf E_i \otimes
\mathbf E_i= \sum_i K_{j}^{ii}\, \mathbf E_i \otimes \mathbf E_i,
\qquad \forall\, j=1,\ldots, n,
\]
where $\mathbf E_i$ are common eigenvectors of $\mathbf K_{j}$ and
$\lambda_{j}^i$ are the corresponding eigenvalues.
\end{proposition}

We call ${\mathbf S}^{-1}=\big( \varphi_{(j)}^i\big)$ the regular
$n\times n$ matrix of the components of $(\mathbf K_{j})$:
  \begin{equation} \label{3.1}
\varphi_{(j)}^i=K_{j}^{ii}= \lambda_{j}^i g^{ii}, \qquad
\varphi_{(n)}^i=K_{n}^{ii}=  g^{ii}.
  \end{equation}
As in the general case (see equation~\eqref{1.4}), its inverse
matrix ${\mathbf S}=\big(\varphi_i^{(j)}\big)$ is a St\"ackel
matrix. Moreover, we consider the invariant $n\times n$ matrix
  $
\Lambda=(\lambda_j^i)
  $
of the eigenvalues of a basis of $\mathcal K_n$, introduced in
Section \ref{s2}. Theorems \ref{t3.6} and \ref{t3.8} below give a
complete algebraic characterization of orthogonal separable webs
in terms of eigenvalues of associated Killing--St\"ackel spaces,
impro\-ving considerably the analysis contained in \cite{[10]} and
providing a rigorous proof of the method illustrated in Section
\ref{s2} for the orthogonal separation.

\begin{proposition}\label{p3.2}
For any fixed index $i=1,\ldots,n$ the fundamental functions
  \begin{equation}
f^{jh}_i=\frac {\det \Lambda_j^i}{\det \Lambda_h^i}, \label{3.2}
  \end{equation}
when well-defined, depend on $q^i$ only: in particular we have
 \begin{equation}
f_i^{jh}=(-1)^{h+j}\frac {\varphi_i^{(j)}}{\varphi_i^{(h)}}.
\label{3.3}
 \end{equation}
\end{proposition}

\begin{proof}
 We have the following
relations between $\mathbf S^{-1}$ and $\Lambda$:
  \begin{equation}
\det \mathbf S^{-1}= \det \Lambda\prod_i g^{ii}, \qquad
\det (\mathbf S^{-1})_h^k=\dfrac{\det
\Lambda_h^k}{g^{kk}} \prod_i g^{ii}. \label{3.4}
  \end{equation}
By def\/inition of inverse matrix and by (\ref{3.4}), we see that
each element of $\mathbf S$ has the following expression
\[
\varphi_i^{(j)}=(-1)^{i+j}\,\frac{\det (\mathbf S^{-1})_j^i}{\det
\mathbf S^{-1}}= (-1)^{i+j}\,\frac{\det
\Lambda_j^i}{g^{ii} \,\det \Lambda}.
\]
Hence, by (\ref{3.2}), equation (\ref{3.3}) holds and the
fundamental functions (\ref{3.2}) depend on the coordinate~$q^i$
only.
\end{proof}

\begin{proposition}\label{p3.3}
For any index $i$ there exist indices $h$ and $j$ with $h<j$ such
that the function~\eqref{3.2} is well-defined.
\end{proposition}

\begin{proof}
Since $\det \Lambda \neq 0$, for each $i$ there exists an index
$h_0$ such that $\det \Lambda^i_{h_0}\neq 0$. If $h_0<n$, then
$f_i^{jh_0}$ is well-def\/ined for any $j>h_0$. If $h_0=n$, then
there exists an index $h<n$ such that $\det \Lambda^i_h\neq 0$.
Indeed, the $n\times (n-1)$ matrix $\Lambda^i$ obtained from
$\Lambda$ by eliminating the $i$-th column has rank $n-1$.
Moreover, being $\det \Lambda^i_n\neq 0$, the f\/irst $n-1$ lines
are independent i.e., form a basis of a $(n-1)$-dimensional linear
space. Since the last row is dif\/ferent from zero (all its
elements are equal to 1), there exists a basis made of the last
row and other $n-2$ rows of $\Lambda^i$. This means that there
exists $h<n$ such that $\det \Lambda^i_h\neq 0$. Hence, for any
index $i$, at least one function $f_i^{jh}$ with $j>h$ is well
def\/ined.
\end{proof}

{}From the Def\/inition \ref{d2.1} of CKV, we get the following
lemma

\begin{lemma} \label{l3.4}  The vector field $\mathbf X=f(q^1,\ldots,q^n)\partial _i$ is a CKV if and only if
(i) $f$ depends on $q^i$ only; (ii) there exists a function $F$
such that
  \begin{equation}
\partial _i\ln g^{jj}=F, \quad j\neq i, \qquad
\partial_i\ln g^{ii}=F+2\partial_i \ln f.
\label{3.5}
  \end{equation}
In particular if $F=0$, then $\mathbf X$ is a KV.
\end{lemma}

\begin{remark}\label{r3.5} \rm
By (\ref{3.5}) it follows that  $\partial_i g^{hh}=\partial_i
g^{jj}$ for all $h,j$ both dif\/ferent from $i$. Moreover, due to
item (i) of Lemma \ref{l3.4}  the coordinate $q^i$ can always be
rescaled in order to have $\mathbf X=\partial_i$.  This means that
if $\partial_i$ is parallel to a CKV (resp. KV), we can assume
without loss of generality that $\partial_i$ is a CKV (resp. KV),
by rescaling the corresponding coordinate.
\end{remark}

\begin{theorem} \label{t3.6}   \it
Let $\mathbf E_i$ be a common eigenvector of a KS-algebra
$\mathcal K_n$. Then,  {\rm (i)} for $n>2$
  $\mathbf E_i$ is proportional to a conformal Killing vector if and only if
the ratios $f_i^{jh}$ are constant or undefined for every $h<j<n$;
 in particular, {\rm (ii)} for any $n$ $\mathbf E_i$ is proportional to a Killing vector if and only if
the ratios $f_i^{jh}$ are constant or undefined for every $h<j$.
\end{theorem}

\begin{proof}  Let $\mathbf E_i$ be a common eigenvector of $\mathcal K_n$ and $(q^i)$ be separable coordinates adapted to $\mathcal K_n$.
For simplicity we take $i=1$. Since the separable coordinates are
orthogonal, $\mathbf E_1$ is proportional to $\partial_1$. Thus,
without loss of generality, we assume that the vector $\mathbf
E_1=\partial_1$ is proportional to a~CKV.

From Proposition \ref{p3.2} we have for all $j\neq 1$
  \begin{equation}
\partial _1f_j^{hk}=0.
\label{3.6}
  \end{equation}
Let us consider $\partial _1f_1^{hk}$. Due to properties of
determinants, we have
  \begin{equation}
\partial _1\det \Lambda ^1_k =\sum _{p=2}^n \Xi_p,
\label{3.7}
  \end{equation}
where $\Xi _p $ is the determinant of the matrix obtained from
$\Lambda_k^1$ by replacing the elements $(\lambda ^p_h)$ of its
$p$-th column  with $(\partial_1\lambda^p_h)$. By equations
(\ref{1.5})$_1$ and (\ref{3.5}), for all $h$, $k$, we have
  \begin{equation}
\partial _1\lambda^h_k=(\lambda^1_k-\lambda^h_k)F.
\label{3.8}
  \end{equation}
By substituting  (\ref{3.8}) in (\ref{3.7}), we obtain
  \begin{equation}
\partial _1\det \Lambda ^1_k=-F\left( (n-2)\det \Lambda ^1_k-\sum _{p=1}^n(-1)^p\det \Lambda_k^p\right).
\label{3.9}
  \end{equation}
We remark that $\sum\limits_{p=1}^n(-1)^p\det \Lambda_k^p$ is (up
to the sign) the determinant of the matrix obtained from~$\Lambda$
by replacing the $k$-th row with the row made by $n$ elements
equal to~1. Moreover, since for $k\neq n$ also the last row of
$\Lambda_k^1$ contains $n$ elements equal to 1, we have
  \begin{equation}
\sum _{p=1}^n(-1)^p\det \Lambda_k^p=0 ,\quad k\neq n,\qquad
 \sum _{p=1}^n(-1)^p\det \Lambda_n^p=(-1)^{n+1}\det \Lambda.
\label{3.10}
  \end{equation}
We can now evaluate $\partial _1f_1^{hk}$ for all indices $j$, $h$
such that  the function is well-def\/ined. Recalling that we
always assume  $k< h$,  from (\ref{3.9}) and (\ref{3.10}) it
follows
  \begin{equation}
\partial_1f_1^{hk}=0,\quad h\neq n,\qquad
\partial_1f_1^{nk}=(-1)^{n+1}F \dfrac {\det \Lambda}{\det \Lambda^1_k}.
\label{3.11}
  \end{equation}
This proves that $f_1^{hk}$ is constant for $k<h<n$. In particular
if $F=0$ (i.e., $\partial _1$ is proportional to a Killing
vector), then all $f_1^{hk}$ are constant or undef\/ined.

Conversely, let us assume that $f_1^{hk}$ is constant or
undef\/ined for every $k<h<n$. By (\ref{3.3}) there exists an
index $j\neq n$ and $n-1$ real constant $c^{hj}$
($h=1,\ldots,n-1$) such that for all $h< n$
  \begin{equation}
\varphi_1^{(h)}=c^{hj}\varphi _1^{(j)} \label{3.12}
  \end{equation}
and, due to the regularity of $\mathbf S$, $\varphi _1^{(j)}\neq
0$. Let $\tilde {\mathbf S}$ be the $n\times n$ matrix obtained
from $\mathbf S$ by dividing the f\/irst row by
$\varphi_{1}^{(j)}$. The following relations link $\mathbf S$ and
$\tilde{\mathbf S}$
\[
\det \mathbf S= \varphi_{1}^{(j)} \det \tilde {\mathbf S}, \qquad
\det \mathbf S_h^k= \varphi_{1}^{(j)} \det \tilde {\mathbf S}_h^k,
\quad h\neq 1,\qquad \det \mathbf S_1^k= \det \tilde{\mathbf
S}_1^k.
\]
The $n$-th element of the f\/irst row of $\tilde {\mathbf S}$ is
the only element of the matrix depending on $q^1$. Thus,
  \begin{equation}
\partial _1\det \tilde {\mathbf S}=(-1)^{n+1}\partial _1\left( \dfrac {\varphi_1^{(n)} }{\varphi_1^{(j)} }\right)\det \tilde {\mathbf S}^n_1,
\label{3.13}
  \end{equation}
and for the same reason we get, up to the sign,
  \begin{alignat}{3}
&\partial _1\det \tilde {\mathbf S}_h^k=\partial _1\left( \dfrac
{\varphi_1^{(n)} }{\varphi_1^{(j)} }\right)\det \big(
\tilde {\mathbf S}_h^k\big)^n_1,\quad&& h\neq 1\  \hbox{and}\  k\neq n,&\nonumber\\
& \partial _1\det \tilde {\mathbf S}_h^k=0, \quad && h=1 \
\hbox{or} \  k=n.& \label{3.14}
  \end{alignat}

From the def\/inition of inverse matrix, we have
\begin{gather*}
g^{11}=\varphi_{(n)}^1=(-1)^{1+n} \dfrac{\det\mathbf S^n_1}{\det
\mathbf S}=(-1)^{1+n}\dfrac{\det\tilde{\mathbf S} ^n_1}
{\varphi_1^{(j)}\det \tilde {\mathbf S}},\cr
g^{hh}=\varphi_{(n)}^h=(-1)^{h+n} \dfrac{\det\mathbf S^n_h}{\det
\mathbf S}=(-1)^{h+n}\dfrac{\det\tilde{\mathbf S}^n_h}{\det \tilde
{\mathbf S}}, \qquad h\neq 1.
\end{gather*}
Thus, we get
  \begin{equation}
\partial _1g^{11}=(-1)^{1+n}\partial _1\bigg( \dfrac{\det\tilde {\mathbf S}^n_1}{ \varphi_1^{(j)}\det \tilde{\mathbf S}}\bigg),
\qquad
\partial _1g^{hh}=(-1)^{h+n}\partial _1\bigg( \dfrac{\det\tilde {\mathbf S}^n_h}{\det \tilde{\mathbf S}}\bigg),
\label{3.15}
  \end{equation}
for $h\neq 1$. Hence, by (\ref{3.14}$)_2$ we have
\[
\partial _1g^{11}=(-1)^{n}\dfrac { (\det\tilde{\mathbf S}^n_1)\partial _1(\varphi_1^{(j)}\det \tilde{\mathbf S} )}{(\varphi_1^{(j)} \det \tilde{\mathbf S} )^2},
\qquad
\partial _1g^{hh}=(-1)^{h+n-1}\dfrac { (\det\tilde{\mathbf S}^n_h)\partial _1(\det \tilde{\mathbf S})}{(\det \tilde{\mathbf S})^2},
\]
and
  \begin{equation}
\partial _1\ln g^{hh}=-\partial _1\ln (\det \tilde{\mathbf S}), \qquad
\partial _1\ln g^{11}=-\partial _1\ln (\det \tilde{\mathbf S})-\partial_1 \ln \varphi_1^{(j)}.
\label{3.16}
  \end{equation}
By Lemma \ref{l3.4} it follows that $\partial _1$ is proportional
to a CKV with $F=-\partial _1\ln (\det \tilde{\mathbf S})$ and
$f=\sqrt{\varphi_1^{j}}$. We remark that $F$ does not depend on
the choice of the element $\varphi^{(j)}_1$ used in the
construction of $ \tilde{\mathbf S}$. Indeed, for any other $j'$
such that $\varphi^{(j')}_1\neq 0$, by (\ref{3.12}) we have
\[
\varphi_1^{(j')}=c^{j'j}\varphi _1^{(j)}
\]
for a suitable constant $c^{j'j}\in {\mathbb R}$ and $\det
\tilde{{\mathbf S}}={c^{j'j}}\det \tilde{{\mathbf S}}'$, where
$\tilde {\mathbf S}'$ is obtained from $\mathbf S$ by dividing the
f\/irst row by $\varphi^{(j')}_1$. Then, the function $F$ does not
change. In the particular case when  all $f_1^{jh}$ are constant,
then (\ref{3.12}) hold for all $h=1,\ldots, n$ and by (\ref{3.13})
it follows that $F=-\partial _1\ln (\det \tilde {\mathbf S})=0$.
Hence, by (\ref{3.15}) we get $\partial_1 g^{hh}=0$ for any $h\neq
1$ and $\partial_1 g^{11}=-\partial_1 \ln \varphi_1^{(j)}$.
According to Lemma~\ref{l3.4} and Remark~\ref{r3.5}, this means
that $\partial_1$ is proportional to a Killing vector and $q^1$ is
ignorable up to a rescaling. In this case, due to (\ref{1.5})$_1$
all the elements of $\mathbf S^{-1}$ and $\mathbf S$ do not depend
on $q^1$.
\end{proof}

\begin{remark}\label{r3.7} \rm
For $n=2$  every eigenvector of a characteristic Killing tensor is
proportional to a~CKV. This fact can be checked directly by
writing $g^{11}$ and $g^{22}$ in terms of a two dimensional
St\"ackel matrix.
\end{remark}

From Proposition \ref{p3.2} and Theorem \ref{t3.6} it follows that

\begin{theorem}\label{t3.8}  \it
Let $\mathbf E_i$ be a common eigenvector of a KS-algebra. For
every $i=1,\dots, n$ one and only one of the following statements
holds: {\rm I)} $\mathbf E_i$ is, up to a scalar factor,  a KV.
{\rm II)} There exist indices $j,h$ such that, in a neighborhood
of any point $P$ where $df_i^{jh}(P)\neq 0$, the equation
 \begin{equation}
f_i^{jh}=\hbox{\rm const} \label{3.17}
  \end{equation}
defines a hypersurface orthogonal to $\mathbf E_i$. \rm
\end{theorem}

\begin{example} \label{e3.9} \rm
Let us consider a {\it L-tensor} $\mathbf L$, that is a conformal
Killing tensor with simple eigenvalues and vanishing Nijenhuis
torsion~\cite{[5]}. It is known that, if the eigenvalues $(u^i)$
of the L-tensor $\mathbf L$ are functionally independent (this
property is not required in our def\/inition according to~\cite{[5]}), then the $(u^i)$ are a separable coordinate system
for the geodesic Hamilton--Jacobi equation. In \cite{tsiga} a
computer algorithm was implemented for constructing of the
separable coordinates associated with a L-tensor compatible with
the potential of a natural Hamiltonian on a proper Riemannian
manifold. We verify the behaviour of the $f_i^{jh}$ for the
KS-algebra generated by $\mathbf L$. Moreover, we see that our
method allows to f\/ind new properties of these systems. We recall
that a symmetric two-tensor f\/ield $\mathbf K$ is said to be a
{\it conformal Killing tensor} (CKT) if there exists a vector
f\/ield $\mathbf C$ such that
  \begin{equation}
[\mathbf K,\mathbf G]=2\mathbf C\odot\mathbf G, \label{3.18}
  \end{equation}
where  $[\cdot ,\cdot ]$ denotes the Lie--Schouten brackets and
$\odot$ the symmetric tensor product. Then, (see \cite{[5],[3]}),
the tensors $(\mathbf K_0,\mathbf K_1,\ldots, \mathbf K_{n-1})$
where
\[
\mathbf K_0=\mathbf G, \qquad \mathbf K_a=\tfrac 1
{a}{\mathrm{tr}}(\mathbf K_{a-1}\mathbf L)\mathbf G- \mathbf
K_{a-1}\mathbf L, \quad a>1
\]
form a basis of a KS-algebra. The matrix $\Lambda$ of eigenvalues
of the $\mathbf K_a$ is
 $
\Lambda=\big(\sigma_a^i(u^1,\ldots, u^n)\big)$,  $a=0,\ldots,
n-1$, $i=1,\ldots, n,
 $
where $\underline u=(u^1,\ldots, u^n)$ are the eigenvalues of
$\mathbf L$ and, for $a>0$, $\sigma_a^i$ are the elementary
symmetric polynomials of degree $a$ in the $n-1$ variables
$(u^1,\ldots, u^{i-1},u^{i+1},\ldots,u^n)$, for $a=0$ we set
$\sigma_0^i=1$. In this case we have
 \begin{equation}
f_i^{jh} = (u^i)^{h-j}. \label{3.19}
 \end{equation}
Indeed, we have that the inverse matrix of $\Lambda$ is
\[
\mathbf A=(A^a_i)=
\left((-1)^a\frac{(u^i)^{n-a-1}}{U'(u^i)}\right),
\]
where $U'(u^i)$ is a suitable function of $u^i$ (see \cite{[3]})
and the fundamental functions satisfy
\[
f^{jh}_i=\frac{\det \Lambda^i_j}{\det\Lambda^i_h}=(-1)^{j-h}\frac
{A^j_i}{A^h_i}.
\]
In particular, as expected, we get
  $
f_i^{j\,j+1}=u^i.
  $
We notice that, due to (\ref{3.19}), for any f\/ixed $i$ either
the fundamental functions are constant for all $j<h$ (if $u^i\in
{\mathbb R}$),  either none of them is constant (if $u^i$ is a
non-constant function). Thus, by Theorem \ref{t3.6}, we obtain the
following theorem, which provides an interesting restriction upon
L-tensors.
\end{example}

\begin{theorem} \label{t3.10} \it
For $n>2$, a L-tensor has no  eigenvector proportional to a proper
conformal Killing vector.
\end{theorem}

\section{Separable webs in Riemannian \\ and pseudo-Riemannian manifolds}\label{s4}

In this section we prove that the method shown in Section \ref{s2}
is ef\/fective also when separable coordinates are not orthogonal
or the metric is not positive def\/inite and isotropic coordinates
may be present. However, for adapting the results of the
orthogonal case, we have to take in account several dif\/ferences
occurring in the general case.
First of all we recall that in the general case we cannot identify
$\mathbf E_i$ and $\partial _i$ as seen in the previous section.
Indeed, for orthogonal separable coordinates the common
eigenvectors of $\mathcal K_n$ are always in the form $\mathbf
E_i=f^i\, \partial _i$ ($i$ not summed) as well as the
corresponding eigenforms $(\mathbf E_i)^\flat=f_idq^i$ are
proportional to $dq^i$ (see Def\/inition \ref{d1.3}). In the
non-orthogonal case, from Def\/inition \ref{d1.3} still
$dq^a=(\mathbf E_a)^\flat$, but by (\ref{1.1})
  \begin{equation}
\mathbf E_{\hat a}=g^{\hat a \hat a}\partial_{\hat a}, \qquad
\mathbf E_{\bar a}=g^{\bar
a\alpha}\partial_\alpha=\varphi_{(m)}^{\bar a}(\theta_{\bar
a}^\alpha \partial_\alpha). \label{4.1}
  \end{equation}
Thus, for all indices $\hat a$ the eigenvectors $\mathbf E_{\hat
a}$ are proportional to $\partial_{\hat a}$, but for isotropic
coordinates  the f\/ields $\mathbf E_{\bar a}$ are not
proportional to $\partial_{\bar a}$. Moreover, by (\ref{4.1})$_2$,
we see that $\mathbf E_{\bar a}$ are proportional to the vectors
$\theta_{\bar a}^\alpha \partial_\alpha$ which are Killing vectors
of the hypersurfaces orthogonal to $\mathbf E_{\bar a}$. If for
all $\alpha$ the functions $\theta_{\bar a}^\alpha$ are constant,
the vectors $\theta_{\bar a}^\alpha \partial_\alpha$ are KV of the
whole manifold. Since $\mathbf E_{\bar a}$ are null vectors, we
can distinguish between essential eigenvectors of type $\hat a$
and $\bar a$ before knowing the separable coordinates. Moreover,
in non-orthogonal coordinates the characterization of the vector
f\/ields $\partial_a$ associated with essential coordinates $q^a$
as CKV or KV is more complicated than Lemma \ref{l3.4}:

\begin{lemma} \label{l4.1}  Let $(g^{ij})$ in standard form \eqref{1.1} in coordinates $(q^i)=(q^a,q^\alpha)$. The vector $\mathbf X=\partial_a$ is a CKV if and only if  there exists a function $F$ such that
\begin{equation}
\begin{array}{l}
\partial _a\ln g^{\hat b \hat b}=F, \\
\partial _ag^{\bar b \alpha}=Fg^{\bar b \alpha},\\
\partial _ag^{\alpha \beta}=Fg^{\alpha \beta}.
\end{array}
\qquad \hat b=1,\ldots, m_1,\quad \bar b=m_1+1,\ldots,m.
\label{4.2}
\end{equation}
If $F=0$ then $\mathbf X$ is a KV.
\end{lemma}

\begin{proof}  From def\/inition of CKV we have, for $\mathbf X=\partial _a$
\[
\partial_a g^{ij}=Fg^{ij},
\]
for all $i,j=1,\ldots,n$. By (\ref{1.1}), observing that $g^{\hat
b \hat b}\neq 0$, we obtain (\ref{4.2}).
\end{proof}

Unlike the orthogonal case, by (\ref{1.2}) and (\ref{4.2}) we can
see that not only the components of the inverse of a St\"ackel
matrix are involved in the def\/inition of CKV, but  also the
functions~$\theta_{\bar a}^{\alpha}$ and~$\eta_a^{\alpha \beta}$.
Since these functions appear in the metric components, we have to
modify Theorem \ref{t3.6} for the non-orthogonal case. In
particular we need to def\/ine a new kind of vector f\/ields
playing the role of KV and CKV in non-orthogonal separable
coordinates.

\begin{definition} \label{d4.2} \rm   Let $(q^i)=(q^a,q^\alpha)$ be standard separable coordinates. The vector $\mathbf X=\partial_a$ is a~{\it conformal St\"ackel symmetry} (CS-symmetry) of the foliation $S^a$ if there exists a function $F$ such that for all  $b=1,\ldots, m$
\begin{equation}
\partial_a\ln \varphi_{(m)}^b=F.
\label{4.3}
\end{equation}
We say that $\mathbf X$ is a {\it St\"ackel symmetry} (S-symmetry)
if it is a CS-symmetry with $F=0$.
\end{definition}

\begin{remark} \label{r4.3} \rm
Due to the regularity of $(g^{ij})$ the $\varphi_{{(m)}}^b$ are
all dif\/ferent from zero. For the same reason, for a given $ \bar
b$, the $\theta _{\bar b}^\alpha$ are not all zero.
\end{remark}

\begin{proposition} \label{p4.4}   Let $\partial _a$ be a CS-symmetry (S-symmetry). Then, $\partial_a$ is a CKV (KV) if and only if $\theta _a^\alpha $ and $\eta_a^{\alpha \beta}$ are constant for every $\alpha$, $\beta $.
\end{proposition}

\begin{proof}  By (\ref{1.2}) and (\ref{4.3}), for a given CS-symmetry $\partial _a$ we have
\begin{gather*}
\partial _a\ln g^{\hat b \hat b}=F,\\
\partial _ag^{\bar b\alpha}=\delta _a^{\bar b} \partial _a\theta_{\bar b}^\alpha \varphi_{(m)}^{\bar b}+Fg^{\bar b \alpha},\\
\partial_ag^{\alpha \beta}=\partial _a\eta _a^{\alpha \beta} \varphi^a_{(m)}+Fg^{\alpha \beta}.
\end{gather*}
Because of (\ref{4.2}) and of $\varphi ^a_{(m)} \neq 0$, $\partial
_a$ is a CKV if and only if $\theta _{\bar a}^\alpha $ and
$\eta_a^{\alpha \beta}$, both functions of the coordinate
corresponding to the lower index only, are constant for every
$\alpha$, $\beta $.
\end{proof}

\begin{remark} \label{r4.5} \rm Proposition \ref{p4.4} shows that CS-symmetries are not coordinate independent objects unless coordinates are orthogonal, in this case they coincide with CKV.
Also for the isotropic coordinates we have that if $\partial_{\bar
a}$ is a KV then,  by Proposition \ref{p4.4} and (\ref{4.1})$_2$,
$\mathbf E_{\bar a}$ is proportional to the KV $\theta_{\bar
a}^\alpha \partial_\alpha$. If no isotropic coordinate occurs,
then $\partial_a$ is a CS-symmetry if and only if it is a~CKV of
each $m$-dimensional submanifold $\{q^\alpha={\rm const}$,
$\alpha=m+1,\ldots,n\}$.
\end{remark}

As in the previous sections, we introduce the $m\times m$ matrix
$\mathbf \Lambda=(\lambda^a_b)$ of the essential eigenvalues of a
basis $(\mathbf K_1,\ldots, \mathbf K_m=\mathbf G)$ of $\mathcal
K_m$ described in Section \ref{s2}. Even if in the construction of
$\mathbf \Lambda$ we do not explicitly distinguish between
eigenvalues $\lambda^{\hat a}$ and $\lambda^{\bar a}$, we will see
in Remark \ref{r4.13} that the distinction is relevant.

\begin{proposition}\label{p4.6}  The vector $\mathbf X=\partial _a$ is a CS-symmetry if and only if
 \begin{equation}
\partial _a\lambda^b_c=(\lambda^a_c-\lambda ^b_c)F \qquad \forall \;b,\,c=1,\ldots, m,
\label{4.4}
 \end{equation}
where $F$ is the function appearing in \eqref{4.3}. The vector
$\mathbf X=\partial _a$ is a S-symmetry if and only if  \rm
\[
\partial_a \lambda^b_c=0 \qquad \forall\; b,\, c=1,\ldots, m.
\]
\end{proposition}

\begin{proof}  It follows directly from Def\/inition \ref{d4.2} and (\ref{1.5})$_1$, recalling that in $(\mathcal K_m)$ there is  always at least one KT with distinct essential eigenvalues.
\end{proof}

\begin{lemma} \label{l4.7}
Let $S=(\varphi ^{(b)}_a)$ be the $m\times m$ St\"ackel matrix
defined by \eqref{1.2} and \eqref{1.4}. Then
 \begin{equation}
\varphi^{a}_{(b)}=\lambda_b^a \varphi^a_{(m)}, \qquad \forall \;
a,b=1,\ldots, m. \label{4.5}
  \end{equation}
\end{lemma}

\begin{proof}  According to (\ref{1.3}) and (\ref{1.4}),
the non-vanishing components of the basis $(\mathbf K_1,\ldots,
\mathbf K_m)$ in standard coordinates have the following
equivalent forms
  \begin{equation}
 K_b^{\hat a\hat a}=\lambda^{\hat a}_b g^{\hat a\hat a}=\varphi^{\hat a}_{(b)},\qquad
 K_b^{\bar a \beta}=\lambda^{\bar a}_b g^{\bar a\beta}
=\theta^{\beta}_{\bar a}\varphi^{\bar a}_{(b)}. \label{4.6}
  \end{equation}
By inserting the expression of the metric (\ref{1.2}) in
(\ref{4.6}), we get
\[
\lambda^{\hat a}_b \varphi^{\hat a}_{(m)}=\varphi^{\hat a}_{(b)},
\qquad \lambda^{\bar a}_b \theta^{\beta}_{\bar a}\varphi^{\bar
a}_{(m)}=\theta^{\beta}_{\bar a}\varphi^{\bar a}_{(b)}.
\]
Hence, due to Remark \ref{r4.3}, relation (\ref{4.5}) holds for
all essential eigenvalues, without distinction between isotropic
and non-isotropic coordinates.
\end{proof}

\begin{remark}\label{r:ult}
By (\ref{4.5}) for the general separation (cf. (\ref{3.4}) for the
orthogonal case), it follows: $\det \mathbf S^{-1}= \det
\Lambda\prod\limits_a \varphi^a_{(m)}$. Therefore, for
nondegenerate metrics we have always $ \det \Lambda\neq 0. $
\end{remark}

We def\/ine for essential coordinates the fundamental functions
$f_a^{bc}$ in terms of minors of $\Lambda$ as described by
(\ref{2.1}). Namely, Propositions \ref{p3.2} and \ref{p3.3} can be
directly restated here as follows

\begin{proposition} \label{p4.8} \it {\rm (i)} If
 \begin{equation}
f^{bc}_a=\frac {\det \Lambda_b^a}{\det
\Lambda_c^a}=(-1)^{b+c}\frac {\varphi_a^{(b)}}{\varphi_a^{(c)}}
\label{4.7}
  \end{equation}
is well-defined, then it depends on $q^a$ only. {\rm (ii)} For any
fixed index $a$ there exist two indices $c<b\leq m$ such that the
fundamental function $f_a^{bc}$ \eqref{4.7} is well-defined.
\end{proposition}

We can now generalize Theorem \ref{t3.6}:

\begin{theorem}\label{t4.9}
Let $q^a$  be an essential coordinate adapted to a separable
Killing algebra $(D_r,\mathcal K_m)$. Then, {\rm (i)} for $m>2$
there exists a rescaling $\breve{q}^{a}=\breve{q}^a(q^a)$ such
that the  associated vector field~$\breve \partial_a$
 is a CS-symmetry if and only if
the functions $f_a^{bc}$ \eqref{4.7} are constant or undefined for
every $c<b< m$.
 In particular, {\rm (ii)} for any $m$, $\breve\partial_a$ is a S-symmetry if and only if
the functions $f_a^{bc}$ are constant or undefined for every
indices $b$, $c$.
\end{theorem}

\begin{proof}  By comparing (\ref{4.5}) and (\ref{3.1}), we see that the relations
between essential components of Killing tensors and the $m\times
m$ St\"ackel matrix $(\varphi ^{(a)}_c)$ are exactly the same as
in the orthogonal case. To prove our thesis,  we follow the proof
of Theorem \ref{t3.6} with some modif\/ications. Let us assume
$a=1$ and that $\breve\partial_1$ is a CS-symmetry. Then,
equations (\ref{4.4}) hold and by calculating $\breve\partial_1
f_1^{bc}$ as in Theorem \ref{t3.6} we get equations
\[
\breve\partial_1f_1^{bc}=0,\quad c<b<m,\qquad
\breve\partial_1f_1^{mc}=(-1)^{m+1}F \dfrac {\det \Lambda}{\det
\Lambda^1_c},
\]
corresponding to (\ref{3.11}). Hence, the fundamental functions
$f_1^{bc}$ are constant or undef\/ined for all $b<c<m$ and, if
$F=0$ (i.e., $\breve\partial_1$ is a S-symmetry) they are all
constant or undef\/ined. Conversely, if $f_1^{bc}$ are constant or
undef\/ined for all $b<c<m$, by repeating the same reasoning of
Theorem \ref{t3.6} we obtain the following equations analogous to
(\ref{3.16})
\[
\partial _1\ln \varphi_{(m)}^b=-\partial _1\ln (\det \tilde{\mathbf S}), \qquad
\partial _1\ln \varphi_{(m)}^1=-\partial _1\ln (\det \tilde{\mathbf S})-\partial_1 \ln \varphi_1^{(c)},
\]
with $b\neq 1$, where $\varphi_1^{(c)}$ is a non-vanishing element
of the f\/irst row of the St\"ackel matrix $\mathbf S$ and
$\tilde{\mathbf S}$ is the matrix obtained from $\mathbf S$ by
dividing the f\/irst row by $\varphi_1^{(c)}$. If $\partial_1
\varphi_1^{(c)}\neq 0$, we can locally rescale $q^1$ as $\breve
q^{1}=\varphi_1^{(c)}(q^1)$ (if $\varphi_1^{(c)}$ is constant we
do not need to rescale and $\breve\partial_1=\partial_1$). Hence,
for all $b=1,\ldots,m$ we get  $\breve\partial _{1}\ln
\varphi_{(m)}^b=-\breve\partial_{1}\ln (\det \tilde{\mathbf S})$
and by (\ref{4.3}) $\breve\partial_{1}$ is a CS-symmetry with
$F=-\breve\partial_{1}\ln (\det \tilde{\mathbf S})$. In
particular, as in the orthogonal case, if all $f_1^{bc}$ are
constant or undef\/ined then $\det \tilde{\mathbf S}$ is
independent of $\breve q^1$, $F=0$, and $\breve\partial_1$ is a
S-symmetry.
\end{proof}

\begin{remark} \label{r4.10}\rm In the previous theorem no distinction is made between coordinates of type $\hat a$
and~$\bar a$. For $m=2$ it is easy to check that every
$\partial_a$ is up to a rescaling a CS-symmetry.
\end{remark}

By Theorem \ref{t4.9}, Theorem \ref{t3.8} can be generalized in
the following way


 \begin{theorem}\label{t4.11}
Let $(q^a)$ be essential coordinates adapted to a separable
Killing algebra $(D_r,\mathcal K_m)$. For every $a=1,\dots, m$ one
and only one of the following statements holds: {\rm I)} there
exists a~re\-sca\-ling $\breve{q}^{a}=\breve{q}^a(q^a)$ such that
the  associated vector field $\breve \partial_a$ is  a S-symmetry.
{\rm II)} There exist indices $b$, $c$ such that, in a
neighborhood of any point where $df_a^{bc}\neq 0$, the equation
\[
f_a^{bc}=\hbox{\rm const}
\]
defines a hypersurface of the foliation $q^a={\rm const}$.
\end{theorem}

\begin{remark}\label{r4.12}  \rm
The vector f\/ield $\partial_a$ is a S-symmetry if and only if the
St\"ackel matrix does not depend on $q^a$. Therefore, unlike the
orthogonal case,  item I) of Theorem \ref{t4.11}  does not provide
a geometric characterization of the f\/ield $\partial_a$, but
merely a property of the St\"ackel matrix with respect to $(q^i)$,
as it is illustrated in Example \ref{e4.14}. On the contrary, item
II) retains the same geometric meaning as in Theorem \ref{t3.8}.
\end{remark}

\begin{remark}\label{r4.13} \rm
Since $\mathbf E_{\hat a}$ is proportional to $\partial_{\hat a}$,
by applying Theorems \ref{t4.9} and \ref{t4.11} to the fundamental
functions $f_{\hat a}^{bc}$ we have that if $\partial_{\hat a}$ is
a CS-symmetry then $\partial_{\hat a}$
 is a common eigenvector of $\mathcal K_m$ orthogonal to $S^{\hat a}$. This is not true for indices $\bar a$, which correspond to isotropic eigenvectors $\mathbf E_{\bar a}$ generating the isotropic distribution $I=\Delta \cap \Delta ^\perp$.
\end{remark}

\begin{example}\label{e4.14} \rm
Let us consider the Euclidean four-dimensional space ${\mathbb
R}_4$. Let $\mathcal S_2$ be the set of two foliations $S^1$ and
$S^2$ described in Cartesian coordinates $(x,y,z,t)$ as
\[
S^1=\bigcup_{k>0}\big\{(x,y,z,t)\in {\mathbb R}^4 \mid
x^2+y^2=k\big\}, \qquad S^2=\bigcup_{h\in \scriptstyle{{\mathbb
R}}}\big\{(x,y,z,t)\in {\mathbb R}^4 \mid t=h\big\}.
\]
Two vectors orthogonal to $S^1$ and $S^2$ respectively are $
\mathbf n_1=x\,\partial_x+y\,\partial_y,$  $\mathbf
n_2=\partial_t. $ Let $D_2$ be the linear space generated by the
vectors
\[
\mathbf X_3=\partial_z, \qquad \mathbf
X_4=y\,\partial_x-x\,\partial_y,
\]
which are commuting Killing vectors tangent to both foliations
$S^a$. The tensor
\[
\mathbf K=\partial_t\otimes\partial_t
\]
 is a  $D_2$-invariant KT. Moreover, $\mathbf E_1=\mathbf n_1$ and $\mathbf E_2=\mathbf n_2$ are eigenvectors of $\mathbf K$ associated with the eigenvalues $\lambda^1=0$ and $\lambda^2=1$ respectively. Hence, according to Def\/inition~\ref{d1.6},
$(\mathcal S_2,D_2,\mathbf K)$ is a separable Killing web. The
tensors $(\mathbf K, \mathbf G)$ form a basis of the KT-space
$\mathcal K_2$. Let us construct the adapted coordinates
$(q^a,q^\alpha)$ and compute the components of the  metric in
these new coordinates. As essential coordinates we choose
$q^1=\sqrt{x^2+y^2}=\rho$ and $q^2=t$. By adding the ignorable
coordinates $(q^\alpha)=(q^3,\,q^4)$ associated with the basis
$(\mathbf X_3, \mathbf X_4)$ and with a section~$\mathcal Z$
orthogonal to the orbits of $D_2$, the coordinate transformation
is def\/ined by
  \begin{equation}
x=\rho \cos(q^4+\theta_0), \qquad y=\rho \sin(q^4+\theta_0),
\qquad z=q^3+z_0, \qquad t=q^2, \label{4.8}
  \end{equation}
where $\theta_0\in (0,\,2\pi)$ and $z_0\in {\mathbb R}$ are the
parameters def\/ining $\mathcal Z$. In these coordinates the
metric is diagonal and the non vanishing components of $\mathbf G$
are
\[
g^{11}=g^{22}=g^{44}=1, \qquad g^{33}=\rho^{-2}.
\]
By choosing a dif\/ferent basis of $D_2$, for instance $\mathbf
X'_3=\mathbf X_3$ and $\mathbf X'_4=\mathbf X_3+\mathbf X_4$, and
leaving $\mathcal Z$ unchanged, we get non-orthogonal ignorable
coordinates $(q'^\alpha)$ given by
\[
x=\rho \cos(q'^4+\theta_0), \qquad y=\rho \sin(q'^4+\theta_0),
\qquad z=q'^3+q'^4+z_0, \qquad t=q^2.
\]
and the metric assumes the standard form
\[
\mathbf G=\begin{pmatrix} 1 & 0 & 0 & 0 \\
0 & 1 & 0 & 0 \\
0 & 0 & 1+\rho^{-2} & -1 \\
0 & 0 & -1 & 1 \\
\end{pmatrix}.
\]
In both cases the $2\times 2$ St\"ackel matrix associated with the
essential separable coordinates and its inverse are
\[
\mathbf S= \begin{pmatrix} -1 & 1 \cr 1 & 0 \end{pmatrix}, \qquad
\mathbf S^{-1}=\begin{pmatrix} \lambda^1 g^{11} & \lambda^2 g^{22}
\cr g^{11} & g^{22} \end{pmatrix}=
\begin{pmatrix} 0 & 1 \cr 1 & 1 \end{pmatrix},
\]
respectively. The matrix  $\Lambda$ of the essential eigenvalues
of $\mathbf K$ and  $\mathbf G$ coincides with the
mat\-rix~$\mathbf S^{-1}$. The method of the eigenvalues does not
provide any coordinate hypersurface because $\mathbf S$ is
constant. Then, according to Theorem \ref{t4.11}, $\partial_1$,
$\partial_2$ are both S-symmetries. However, from a geometric
point of view the corresponding eigenvectors $\mathbf
E_1=x\,\partial_x+y\,\partial_y$ and $\mathbf E_2=\partial_t$ have
dif\/ferent properties. Indeed, $\mathbf E_2$ is a KV, according
to the fact that $q^2$ is ignorable, while $\mathbf E_1$ is not a
KV since $g^{33}$ depends on $q^1$. We can apply the eigenvalue
method to the orthogonal system~(\ref{4.8}) for computing the
equation of the hypersurfaces of $S^1$. In order to determine the
$4\times 4$ matrix $\Lambda$ we consider the $4$-dimensional KT
space containing $\mathcal K_2$ and the tensors $\mathbf
K_2=\mathbf X_3\otimes\mathbf X_3$, $\mathbf K_3=\mathbf
X_4\otimes\mathbf X_4$, which is a KS-algebra for the orthogonal
system (\ref{4.8}).
\end{example}

\begin{example}
Let us consider a 4-dimensional pseudo-Riemannian manifold having
in a coordinate system $(X,Y,Z,U)$ the following non-zero
contravariant metric  components
\begin{gather*}
g^{11}=-X^{10}\frac{-4(X^4-Y^4)+9X^4Y^{24}\Psi}{144(X^{10}+Y^{10})^2(X^4-Y^4)},
\\
g^{22}=-Y^{10}\frac{-4(X^4-Y^4)-9X^{24}Y^4\Psi}{144X^{10}(X^{10}+Y^{10})^2(X^4-Y^4)},
\\
g^{12}=X^{5}Y^5\frac{4(X^4-Y^4)+9X^{14}Y^{14}\Psi}{144(X^{10}+Y^{10})^2(X^4-Y^4)},
\\
g^{33}=-(Z-U)^2-\Psi, \qquad g^{44}=(Z-U)^2-\Psi, \qquad g^{34}=
-\Psi,
\end{gather*}
where $\Psi=(U-Z+X^6+Y^6)$. For $|Y|<|X|$ and $\Psi>0$ the
signature is (3,1), while for $|Y|>|X|$ and $\Psi<0$ the signature
is (2,2). The 2-dimensional space of the KVs is generated  by
$\mathbf X_1 =\partial_Z+\partial_U$, $\mathbf X_2=\sqrt{2|f|}(X^5
\partial_X-Y^5 \partial_Y),$ where
\[
f=\frac{Y^{14}X^{14}}{32(X^{10}+Y^{10})^2(X^4-Y^4)}.
\]
We call $D_1$ the KV space generated by $\mathbf X_1$. We have
$r=1=m_0$, since $\mathbf X_1$ is an isotropic vector. Let us
consider the independent tensors $\mathbf K_1$, $\mathbf K_2$
whose non null contravariant components are
\begin{gather*}
K_1^{11}=Y^{10}(U-Z)f, \qquad K_1^{22}= X^{10}(U-Z)f,  \qquad
K_1^{12}= -X^5-Y^5 (U-Z)f,
\\
K_1^{33} =1+ \tfrac 12 (Z-U)^2- \tfrac 12 (Z-U), \qquad K_1^{34}
=1- \tfrac 12 (Z-U), \\ K_1^{44} = 1-\tfrac 12 (Z-U)^2- \tfrac 12
(Z-U),
\\
K_2^{11}=2Y^{10}f, \qquad K_2^{22}= 2X^{10}f,  \qquad K_2^{12}=
-2X^5Y^5 f, \qquad K_2^{33}=K_2^{34}=K_2^{44}=1.
\end{gather*}
The space $\mathcal K_3$ generated by $(\mathbf K_1, \mathbf
K_2,\mathbf G)$ satisf\/ies the hypotheses of Theorem \ref{t1.2}.
Thus, we can apply the eigenvalue method to get the  equations of
the separated coordinate hypersurfaces. The matrix of the
eigenvalues is
\[
\Lambda =
\begin{pmatrix}
0 & \frac{Z-U}{2\Psi}& -\frac 12  \\
0 & -\frac 1\Psi & 0  \\
1 & 1 & 1 \\
\end{pmatrix}.
\]
We get
\[
f_1^{23}=-X^6-Y^6, \qquad f_2^{ab} \quad \hbox{constant or n.d.},
\qquad f_3^{21}=\tfrac 12 (U-Z).
\]
This means that
 \[
 x = X^6 +Y^6, \qquad z = (Z-U)/2
  \]
are essential separable coordinates (the last one is a null
coordinate) and that, up to a rescaling, the vector $ \partial_y$
associated with the remaining essential separable coordinate $y$
is a S-symmetry. The separable coordinate $y=1/Y^4-1/X^4$  cannot
be computed by the eigenvalue method. The coordinate associated
with $\mathbf X_1 \in D_1$ is $u=(U+Z)/2$. By performing the
change of variables $(X,Y,Z,U)\to (x,y,z,u)$ we get the metric in
standard form
\[
\mathbf G =
\begin{pmatrix}
1& 0& 0 & 0 \\
0 & \frac{2x-z}{y}& 0 & 0  \\
0 &   0&  0& -2z^2  \\
0 & 0 & -2z^2& 2z-x \\
\end{pmatrix}.
\]
In the separable coordinates the general KV is $c_1\frac {1}
{\sqrt y}\partial_y + c_2\partial_u$ and the tensors $\mathbf
K_1$, $\mathbf K_2$ become
\[
\mathbf K_1=-\frac zy \partial_y \odot \partial_y +
2z^2\partial_z\odot\partial_u +(1-z)\partial_u\odot\partial_u,
\qquad \mathbf K_2= \frac 1y \partial_y \odot \partial_y +
\partial_u\odot\partial_u.
\]
We remark that $\mathbf K_2$ is a reducible KT (i.e., sum of
symmetric products of KV), while $\mathbf K_1$ is an irreducible
tensor.
\end{example}

\section{Conformal separable orthogonal systems.}\label{s5}

The method developed in Section \ref{s2} characterizes also
conformal separable orthogonal webs \cite{[7]} in a natural way.
We recall that


\begin{definition} \label{d5.1} \rm
The geodesic Hamiltonian $G$ is {\it conformal separable} if there
exists a function $\sigma$ on $Q$ such that the conformal geodesic
Hamiltonian $\bar G=G/\sigma$ (associated with the conformal
metric $\bar{\mathbf G}=\mathbf G/\sigma$) is separable. We call
{\it conformal separable} the coordinates $(q^i)$ allowing the
separation of $\bar G$.
\end{definition}

\begin{remark}
An important application of conformal separable coordinates is the
fact that coordinates allowing $R$-separation of the Laplace
equation are necessarily conformal separable (see
\cite{[17],[11]}). Moreover, in conformally f\/lat manifolds, all
the conformal separable coordinates are also $R$-separable (see
\cite{[18]}).
\end{remark}

Due to Def\/inition \ref{d5.1}, the conformal separation in
orthogonal coordinates is equivalent to the existence of a
KS-algebra for a conformal metric $\bar {\mathbf G}$. The
following theorem contains an intrinsic characterization in terms
of the original metric tensor $\mathbf G$, involving conformal
Killing tensors (CKT), introduced in Example \ref{e3.9}.

\begin{theorem}[\cite{[7]}] \label{t5.2}
The geodesic Hamiltonian $G$ is conformal separable in orthogonal
coordinates if and only if there exist $n$ CKT $(\mathbf K_i)$
pointwise independent with common eigenvectors $(\mathbf E_i)$ and
in conformal involution (i.e., there exist vector fields $\mathbf
C_{ij}$ such that $[\mathbf K_i,\mathbf K_j]=\mathbf C_{ij}\odot
\mathbf G$). It is not restrictive to assume $\mathbf K_n=\mathbf
G$. Each conformal separable coordinate hypersurface is orthogonal
to one of the $n$ common normal eigenvectors of  $(\mathbf K_i)$.
\end{theorem}

We have \cite{[14],[7]}

\begin{proposition} \label{p5.3}  Let $(\mathbf K_1,\ldots,\mathbf K_n=\mathbf G)$  be a set of independent CKT associated with conformal separable orthogonal coordinates $(q^i)$, and let $(\lambda_i^j)$ be their eigenvalues with respect to the metric~$\mathbf G$. Then, for any choice of the index $k=1,\ldots,n$, the tensors
  \begin{equation}
\bar {\mathbf K}_i=\mathbf K_i-\lambda_i^k\mathbf G \qquad
i=1,\dots, n-1 \label{5.1}
  \end{equation}
are KT for the metric
 \begin{equation}
\bar {\mathbf K}_n=\bar {\mathbf G}=(g^{kk})^{-1}\mathbf G
\label{5.2}
  \end{equation}
and $(\bar {\mathbf K}_1,\ldots,\bar {\mathbf K}_n=\bar {\mathbf
G})$ is a basis for the KS-algebra associated with $(q^i)$.
\end{proposition}

\begin{remark}\label{r5.4} \rm
We say that $(\mathbf K_1, \dots ,\mathbf K_{n-1},\mathbf G)$ is a
basis of the {\it conformal Killing space} (CK-space) associated
with the conformal separable coordinates $(q^i)$.
\end{remark}

\begin{sloppypar}
Due to Proposition \ref{p5.3}, by considering the matrix $\bar
\Lambda$ made by the eigenvalues of $(\bar {\mathbf K}_i)$
 with respect to $\bar {\mathbf G}$, we can apply Theorems \ref{t3.6}, \ref{t3.8} for characterizing any orthogonal conformal separable  web associated with $(\mathbf K_1, \dots ,\mathbf K_{n-1},\mathbf G)$ as orthogonal separable web associated with the KS-algebra generated by $(\bar {\mathbf K}_1, \dots ,\bar {\mathbf K}_{n-1},\bar {\mathbf G})$.
Following Section \ref{s2}, we def\/ine
  \begin{equation}
\bar f_i^{jh}=(-1)^{j+h}\dfrac {\det \bar \Lambda_j^i}{\det \bar
\Lambda_h^i}, \label{5.3}
  \end{equation}
 where the matrix $\bar \Lambda$ is formed by the eigenvalues of $(\bar {\mathbf K}_i)$
 with respect to $\bar {\mathbf G}$, and
  \begin{equation}
f_i^{jh}=(-1)^{j+h}\dfrac {\det \Lambda_j^i}{\det \Lambda_h^i},
\label{5.4}
  \end{equation}
where $\Lambda$ is the matrix made by eigenvalues of $(\mathbf
K_i)$ with respect to $\mathbf G$. We remark that
functions~(\ref{5.3}) are not intrinsically def\/ined, since to
determine the $\bar{\mathbf G}$-eigenvalues of the tensors $(\bar
{\mathbf K}_i)$ it is necessary to know the coordinates because of
the factor $g^{kk}$ appearing in (\ref{5.2}). On the contrary, in
functions (\ref{5.4}) only the eigenvalues of tensors satisfying
intrinsic conditions (the hypotheses of Theorem \ref{t5.2}) are
involved.
\end{sloppypar}

\begin{remark} \label{r5.5} \rm Def\/inition \ref{d5.1} of  ($\bar {\mathbf K}_i$) implies that in $\bar \Lambda$ the $k$-th column has $n-1$ zeros.
\end{remark}

\begin{proposition} \label{p5.6} Let $\bar f_i^{jh}$ and $f_i^{jh}$ be the functions defined in \eqref{5.3} and \eqref{5.4} respectively.
Then, for $h<j< n$ $(n>2)$ we have either
\[
\bar f_i^{jh}=f_i^{jh},
\]
or both functions are undefined.
\end{proposition}

\begin{proof}
The eigenvalues of $\bar {\mathbf K}_i$ with respect to $\bar
{\mathbf G}$ are
\begin{equation}
\bar \lambda_i^j=(\lambda_i^j-\lambda_i^k)g^{kk}. \label{5.5}
\end{equation}
By linear algebra arguments, we have that
\[
\det \bar \Lambda_{h}^i=\big(g^{kk})^{n-2} \det \Lambda_{h}^i.
\]
Hence, by (\ref{5.3}) and (\ref{5.4}) the thesis follows.
\end{proof}


\begin{remark} \label{r5.7} \rm A vector f\/ield $\mathbf X$ is CKV for $\mathbf G$ if and only if it is a CKV for any metric conformal to $\mathbf G$.
Thus, in the following for CKV we shall not specify which is the
metric tensor considered.
\end{remark}

In the case of the orthogonal conformal separation Theorems
\ref{t3.6} and \ref{t3.8}  can be restated as follows

\begin{theorem} \label{t5.8}
Let  $\mathbf E_i$ be a common eigenvector of a basis $(\mathbf
K_i)$ of a CK-space. Then, $\mathbf E_i$ is proportional to  a CKV
if and only if for all $h<j<n$ $(n>2)$ the functions $f_i^{jh}$
defined by \eqref{5.4} are constant or undefined.
\end{theorem}

\begin{proof}
Due to Proposition \ref{p5.6} and Remark \ref{r5.7}, the thesis
follows by applying Theorem \ref{t3.6} (i) to the KS-algebra
(\ref{5.1}), (\ref{5.2}) with $k\neq i$.
\end{proof}

\begin{theorem} \label{t5.9}  Let $\mathbf E_i$ be a common eigenvector of a basis $(\mathbf K_i)$ of a CK-space. For every $i=1,\dots, n$  $(n>2)$ one and only one of the following statements holds:
{\rm I)} $\mathbf E_i$ is, up to a scalar factor,  a CKV. {\rm
II)} There exist indices $h<j<n$ such that, in a neighborhood of
any point where $df_i^{jh}\neq 0$, the equation
\[
f_i^{jh}=\hbox{\rm const}
\]
defines a hypersurface orthogonal to $\mathbf E_i$.
\end{theorem}

In the case of conformal orthogonal separation, we do not
distinguish if $\mathbf E_i$ is proportional to a CKV or a KV. The
following property holds

\begin{proposition} \label{p5.10}
 If $\mathbf E_i$ is, up to a factor, a CKV, then it is a KV of $\bar {\mathbf G}=\mathbf G/g^{kk}$ for any $k\neq i$.
\end{proposition}

\begin{proof}
By (\ref{5.1}), we get a basis of the KS-algebra with respect
to $\mathbf G/g^{kk}$ with $k\neq i$. According to Remark
\ref{r5.5}, the $k$-th column of $\bar \Lambda$ has $n-1$ zeros.
Therefore, all submatrices of kind $\bar \Lambda_n^h$, $h\neq k$
have null determinants. This means that   for all $h<n$ the
functions $\bar f_i^{hn}$ are undef\/ined or identically null.
Moreover, since according to Remark \ref{r5.7}  $\mathbf E_i$ is
up to a factor a CKV for $\bar {\mathbf G}=\mathbf G/g^{kk}$, due
to Theorem \ref{t3.6} (i) we get that for $h<j<n$ the functions
$\bar f_i^{hj}= f_i^{hj}$ are constant or undef\/ined. Then,
Theorem \ref{t3.6} (ii) implies that $\mathbf E_i$ is a KV for
$\bar {\mathbf G}$.
\end{proof}

\begin{example} \label{e5.11} \rm
Let us consider in ${\mathbb R}^3$  the vector f\/ields $\mathbf
R_3$ and $\mathbf I_3$ having Cartesian components
\[
\mathbf R_3=(-y,x,0),\qquad \mathbf I_3=\big(-2xz,\,-2yz,\,
x^2+y^2-z^2\big),
\]
respectively. The vector $\mathbf I_3$ is a CKV with respect to
the Euclidean metric $\mathbf G$: it is the inversion with respect
to a generic point on the axis $z$. The vector $\mathbf R_3$ is
the rotation around the $z$ axis and it is a Killing vector. It is
straightforward to check that the two vector f\/ields commute so
that the corresponding linear f\/irst integrals are in involution.
Moreover, the tensors $(\mathbf K_1=\mathbf I_3\otimes \mathbf
I_3,\, \mathbf K_2= \mathbf R_3\otimes \mathbf R_3, \mathbf G)$
are pointwise independent. Hence, they satisfy the hypotheses of
Theorem \ref{t5.2} and they are associated with some conformal
separable coordinate system. We apply the above described method
to determine the conformal separable coordinate hypersurfaces.
Being $\mathbf R_3\perp \mathbf I_3$, the common eigenvectors are
\[
\mathbf E_1=\mathbf I_3\times \mathbf R_3,\qquad \mathbf
E_2=\mathbf I_3,\qquad \mathbf E_3=\mathbf R_3.
\]
The eigenvalues matrix $\Lambda$ is
\[
\Lambda=\begin{pmatrix}
0 & \mathbf I_3\cdot \mathbf I_3 & 0 \\
0 & 0& {\mathbf R_3\cdot \mathbf R_3} \\
1 & 1 & 1
\end{pmatrix}.
\]
The coordinate hypersurfaces orthogonal to $\mathbf E_1$ are the
level sets of the function
\[
f_1^{21}=\dfrac{\left|\begin{matrix}
{\mathbf I_3\cdot \mathbf I_3} & 0 \\
 1 & 1 \\
\end{matrix}\right|
}{\left|\begin{matrix}
0& {\mathbf R_3\cdot \mathbf R_3} \\
 1 & 1 \\
\end{matrix}\right|
}=\dfrac{\mathbf I_3\cdot \mathbf I_3}{-\mathbf R_3\cdot \mathbf
R_3}=-\dfrac{(x^2+y^2+z^2)^2}{x^2+y^2}
\]
which describes the rotational surface obtained by rotating around
the $z$-axis a circle in the plane $(x,z)$ tangent in the origin
$O$ to the $z$-axis (toroids without center opening). Since
$f_1^{12}$ is not constant and both the upper indices are
dif\/ferent from 3, the eigenvector $\mathbf E_1$ is not
proportional to a CKV. According to the fact that $\mathbf I_3$
and $\mathbf R_3$ are conformal Killing tensors, all functions
$f_i^{jh}$ for $i=2,3$ and $h,j\neq 3$ are constant or
undef\/ined. It is well known that the surfaces orthogonal to
$\mathbf E_3=\mathbf R_3$ are half-planes issued from the
$z$-axis. Moreover, it is easy to check that the spheres tangent
in $O$ to the $xy$-plane are hypersurfaces orthogonal to $\mathbf
E_2=\mathbf I_3$. Indeed, the gradient of the function
$q^2=(x^2+y^2+z^2)/z$ is up to the factor $z^2$ exactly $\mathbf
I_3$. The coordinates associated with $(\mathbf K_1, \mathbf K_2,
\mathbf G)$ are known as tangent-spheres coordinates \cite{[15]}
related to $(x,y,z)$ by
\[
x=\frac{\mu \cos \psi}{\mu^2+\nu^2}, \qquad y=\frac{\mu \sin
\psi}{\mu^2+\nu^2}, \qquad z=\frac{\nu}{\mu^2+\nu^2},
\]
where $q^1=\mu$, $q^2=\nu$, $q^3=\psi$ (see also \cite{[16]} for a
detailed analysis and classif\/ication of the symmetric conformal
separable coordinates in $\mathbb R^3$ and the associated CKTs). A
conformal metric which is separable in these coordinates is for
instance $\bar {\mathbf G}=(\mathbf R_3\cdot \mathbf R_3)\,\mathbf
G$. By Proposition~\ref{p5.3}, the tensors $\mathbf K_1$ and
$\mathbf K_2$ are KT for $\bar {\mathbf G}$. By
Proposition~\ref{p5.10}, $\mathbf E_2$ is a KV  for $\bar {\mathbf
G}$. By def\/inition of~$\mathbf K_2$ and because it is a KT for
$\bar {\mathbf G}$, $\mathbf E_3$ also is a KV for $\bar {\mathbf
G}$.
\end{example}

\section{Conclusion}

By using simple arguments of linear algebra and the properties of
the St\"ackel matrices, we have seen how to construct  separable
hypersurfaces by means of eigenvalues of symmetric two-tensors in
Riemannian and pseudo-Riemannian manifolds. It follows that the
webs associated with these hypersurfaces have the same domain of
def\/inition of the eigenvalues employed in the costruction, apart
some closed singular set where the common eigenspaces of the
tensors in the KS (CKS) spaces are not one-dimensional. In (real)
pseudo-Riemannian manifolds, KTs (and CKTs) may have complex
conjugated eigenvalues, in this case it is not possible to
def\/ine real separable coordinates. However, it is possible to
introduce separated complex variables allowing the Jacobi
integration (see \cite{[12]}). The application of our eigenvalue
method to the complex case is in progress \cite{mink}. For
manifolds of constant curvature the whole spaces of Killing and
conformal-Killing tensors are well known, then it is possible to
apply our method to get computer-graphical representations of the
webs. We remark that the separable (resp. conformal-separable)
coordinates here considered are the only ones allowing separation
(resp.\ Fixed Energy $R$-separation \cite{[11]}) of Laplace,
Helmholtz and Schr\"odinger equations \cite{[6]}.

\subsection*{Acknowledgements}

This research is partially supported by MIUR (National Research
Project ``Geometry of Dynamical System'') and by the research
project ``Progetto Lagrange" of Fondazione CRT.

\pdfbookmark[1]{References}{ref}
\LastPageEnding

\end{document}